\documentclass[11pt]{article}
\usepackage{teschool,epsfig,amssymb}
\usepackage{graphicx}
\usepackage[dvips,usenames]{color}
\begin{document}
\vspace*{4cm}
\title{
 Heavy Flavor Probes of Quark-Gluon Plasma}
\author{
 A. A. Isayev}
 \address{Kharkov Institute of
Physics and Technology, Academicheskaya Str. 1, Kharkov, 61108,
Ukraine\\Kharkov National University, Svobody Sq., 4, Kharkov,
61077, Ukraine} \maketitle\abstracts{Some aspects of heavy flavor
probes of quark-gluon plasma (QGP) including quarkonia and open
heavy flavor
 D- and B-mesons, aimed to study the properties of QGP, are discussed in this lecture note.}

\section{Introduction}

 Lattice QCD (LQCD) calculations predict that at a critical temperature
  $T_c\simeq 170\,\mbox{MeV}$, corresponding to an
energy density  $\varepsilon_c \simeq 1\,\mathrm{GeV/fm^3}$,
nuclear matter undergoes a phase transition to a deconfined state
of quarks and gluons, called Quark-Gluon Plasma (QGP). At the
modern collider facilities such as CERN supersynchrotron (SPS)
(the nucleon-nucleon (NN) centre-of-mass energy for collisions of
the heaviest ions  is $\sqrt{s} =17.3\,\mbox{GeV}$), Relativistic
Heavy Ion Collider (RHIC) at Brookhaven ($\sqrt{s}
=200\,\mbox{GeV}$), and Large Hadron Collider (LHC) at CERN
($\sqrt{s} =5.5\,\mbox{TeV}$), whose heavy-ion program  will start
soon, heavy-ion collisions are used to attain the energy density,
exceeding $\varepsilon_c$. This makes the QCD phase transition
potentially realizable within the reach of the laboratory
experiments. The objective is then to identify and to assess
suitable QGP signatures, allowing to study the properties of QGP.
To that end, a variety of observables (probes) can be
used~\cite{Isayev:JPG1,Isayev:JPG2}. Further we will be mainly
interested in heavy-flavour probes of QGP, i.e., utilizing the
particles having c- and b-quarks.

A special role of heavy $Q = (c, b)$ quarks as probes of the
medium created in heavy-ion collision (HIC) resides on the fact
that their masses ($m_c\approx 1.3\,\mbox{GeV},\, m_b\approx
4.2\,\mbox{GeV}$) are significantly larger than the typically
attained ambient temperatures or other nonperturbative scales,
$m_Q\gg T_c, \Lambda_{QCD}=0.2\,\mbox{GeV}$~\cite{Isayev:RH}. This
has several implications: (i) The production of heavy quarks is
essentially constrained to the early, primordial stages of  HIC.
Hence, heavy quarks can probe the properties of the dense matter
produced early in the collision. (ii)~Thermalization of heavy
quarks is "delayed" relative to light quarks. One could expect
that heavy quarks could "thermalize" to a certain extent, but not
fully on a timescale of the lifetime of the QGP. Therefore, their
spectra could be significantly modified, but still would retain
memory about their interaction history, and, hence, represent a
valuable probe. (iii) RHIC, and especially LHC experiments allow
to reach very low parton momentum fractions $x$, where gluon
saturation effects become important. Heavy quarks are useful tools
to study gluon saturation, since, due to their large masses, charm
and bottom cross sections are calculable via perturbative QCD  and
their yield is sensitive to the initial gluon density.

The heavy-flavor hadrons we will be interested in include: 1) open
charm $D=(c \bar q)$ and open beauty $B=(b \bar q)$ mesons
composed of a heavy quark $Q=(c, b)$ and a light antiquark $\bar
q=(\bar u, \bar d)$. These mesons could be sensitive to the energy
density of the medium through the mechanism of in-medium energy
loss; 2)  hidden charm [charmonia=($c \bar c$)] and hidden beauty
[bottomonia=($b \bar b$)] mesons (called collectively heavy
quarkonia) being the bound states of the charm quark-antiquark, or
bottom quark-antiquark pairs, respectively. Heavy quarkonia could
be sensitive to the initial temperature of the system through the
dissociation due to color screening of the color charge that will
be discussed later.

  For detecting heavy flavor hadrons, different decay channels
are used. At LHC, the ALICE experiment for detecting open charm
will use  the reconstructed hadronic decays like $D^0\rightarrow
K^-\pi^+$, $D^+\rightarrow K^-\pi^+\pi^+$, etc., for detecting
open beauty ALICE will use inclusive single lepton decays
$B\rightarrow  e+X$ (midrapidity), $B\rightarrow \mu+X $ (forward
rapidity) and inclusive displaced charmonia decays $B\rightarrow
J/\psi (\rightarrow e^+e^-)+X$. At RHIC, in PHENIX and STAR
experiments  the measurement of the spectra of open heavy flavors
is based on the measurement of the spectra of heavy flavor (HF)
electrons and positrons [($e^++e^-)/2$] from the semileptonic
decays like $D^0\rightarrow K^-e^+\nu_e$, $D^+\rightarrow \bar K^0
e^+ \nu_e$, etc. These measurements are based on the fact that the
decay kinematics of HF electrons/positrons largely conserves the
spectral properties of the parent particles. Besides, the STAR
experiment has the capability to directly reconstruct open heavy
flavor mesons through the hadronic decay channels ($D^0\rightarrow
K\pi$, $D^\pm\rightarrow K\pi\pi$, etc.). At low $p_T$, the STAR
experiment also uses heavy flavor decay muons to provide open
heavy flavor measurements. In all of the above experiments,
quarkonia are detected through their dilepton decays $Q\bar
Q\rightarrow e^+e^-$ (midrapidity), $Q\bar Q\rightarrow
\mu^+\mu^-$ (forward rapidity).

\section{Heavy Flavor Probes of QGP: Quarkonia}

\begin{figure}[tb]\centering
\includegraphics[height=5.5cm,keepaspectratio]{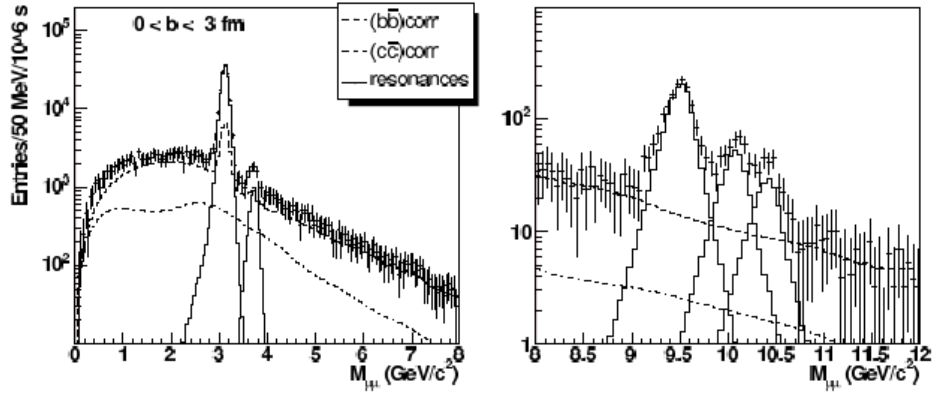}
\caption{Left: Dimuon invariant mass spectra (after subtraction of
the uncorrelated background) in central Pb-Pb collisions at
LHC/ALICE for one year of data taking ($10^6$ s) and luminosity
$5\cdot 10^{26}\,\mathrm{cm}^{-2}\mathrm{s}^{-1}$ in the  $J/\psi$
mass region (PYTHIA calculations). Right:~Same as  on the left,
but in the $\Upsilon$ mass region.
 } \label{Isayev:fig:1}
\end{figure}

We begin the discussion of heavy flavor probes of QGP with heavy
quarkonia. The question we would like to address is: What could
happen with quarkonium yields in HIC  if QGP is really formed? In
the above mensioned experiments, quarkonia are detected in the
dielectron channel at midrapidity and in the dimuon channel at
forward rapidity. For example, in the dimuon channel, signals of
about $7\cdot 10^5$ $J/\psi$s and $10^4$ $\Upsilon$s are expected
in the most central Pb-Pb collisions at LHC/ALICE for one year of
data taking at nominal luminosity (see Fig.~\ref{Isayev:fig:1}
from~\cite{Isayev:JPG2}). If quarkonia will be completely
suppressed by QGP, these signals will disappear, otherwise, in the
case of enhancement of quarkonium yields, the peaks will be more
pronounced as compared to the signals obtained in $pp$ collisions
after scaling them with the corresponding number of independent
binary NN  collisions.

Let us note that if the deconfinement phase transition really
takes place in a dense medium created in  HIC, then a color charge
in the QGP will be screened analogously to the Debye screening of
an electric charge in the electromagnetic plasma. As a result of
color Debye screening of the heavy quark interaction in QGP, the
binding energy of a bound state decreases and one can expect that
this would lead to the suppression of quarkonium yields in HIC.
This idea was first suggested by H. Matsui and
H.~Satz~\cite{Isayev:MS} who predicted that color Debye screening
will result in the suppression of $J/\psi$ meson ($c\bar c$ in the
$^3S_1$ state, $M=3.097\, \mathrm{GeV}$) yields. After that, the
$J/\psi$ suppression  was considered as one of the key probes for
the QGP formation in heavy ion collisions. $J/\psi$ is especially
promising because of the large production cross-section and
dilepton decay channels which make it easily detectable.

However, soon it was realized that besides melting $J/\psi$ mesons
in QGP due to the screening of the color charge, there are also a
few competing mechanisms which could explain the suppression of
$J/\psi$ production in heavy-ion collisions. These mechanisms are
referred to as cold nuclear matter (CNM) effects. The first CNM
 effect is the absorption of $J/\psi$ by
nuclear fragments from colliding nuclei. Let us consider, e.g.,
the proton-nucleus collision. Once produced in the hard primary
parton processes, $J/\psi$ has to cross the length $L$ of nuclear
matter, before exiting the nucleus, and, when traversing nuclear
matter, it can be absorbed by forthcoming nucleons of a nucleus.
The production cross section of $J/\psi$ is $p$-$A$ collision can
be parameterized as \begin{equation}
\sigma_{pA}^{J/\psi}=A\sigma_{pp}^{J/\psi}e^{-\sigma_{abs}^{J/\psi}\varrho
L},
\end{equation}
 where $\sigma_{pp}^{J/\psi}$ is the production cross section of
$J/\psi$ in $pp$ collisions and $\sigma_{abs}^{J/\psi}$ is the
nuclear absorption cross section. From  the global fit to the data
on charmonium  production in $p$-$A$ collisions the value of
$\sigma_{abs}^{J/\psi}$ can be extracted, in particular, at SPS
(NA50 experiment) it was obtained that $\sigma_{abs}^{J/\psi}=
4.2\pm 0.5$ mb~\cite{Isayev:Al}.

The second CNM effect is related to shadowing of low momentum
partons. This means the depletion of low momentum partons in
nucleons bound in nuclei as compared to free nucleons. This effect
can be accounted for in terms of the modification of the parton
distribution functions in nucleon within the nucleus with respect
to the parton distribution functions in a free nucleon:
\begin{equation} R_i^A(x,Q^2)=\frac{f_i^A(x,Q^2)}{
f_i^N(x,Q^2)}<1,\quad i=q_v,q_{sea},g.\end{equation}  Here "i"
denotes valence quarks, sea quarks, and gluons, $x$ is the parton
momentum fraction, $Q^2$ is the momentum transfer squared. At high
energies, $J/\psi$s are dominantly produced through the gluon
fusion,  and the $J/\psi$ yield is therefore sensitive to gluon
shadowing. The underlying idea explaining the occurrence  of gluon
shadowing  is that the gluon density strongly rises at small $x$
to the point where gluon fusion, $gg\rightarrow g$, becomes
significant. In the case of proton-nucleus and nucleus-nucleus
collisions, where nuclei with large mass number $A$ are involved,
the nonlinear effects are enhanced by the larger density of gluons
per unit transverse area of the colliding nuclei. A direct
consequence of nuclear shadowing is the reduction of
hard-scattering cross sections in the phase-space region
characterized by small-$x$ incoming partons. For gluons, e.g.,
shadowing becomes important at $x\lesssim5\times10^{-2}$, and,
hence, is relevant for the conditions of RHIC and LHC. Note
however that the strength of the reduction is constrained by the
current experimental data only for $x\gtrsim10^{-3}$.

\begin{figure}[tb]\begin{center}
\includegraphics[width=6.8cm,keepaspectratio]{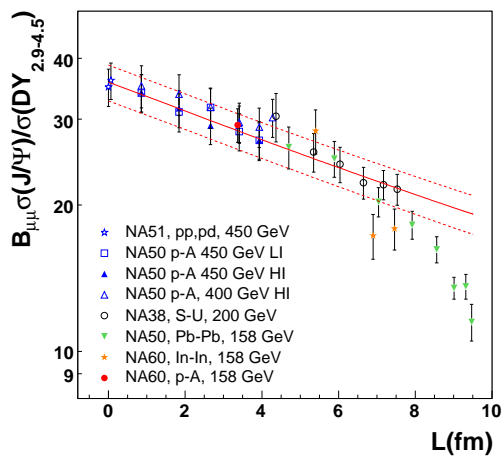}
\end{center} \caption{$J/\psi$ yields normalized by Drell-Yan
process, as a function of the nuclear thickness $L$, as measured
at the SPS.
 } \label{Isayev:fig:3}
\end{figure}

After discussing CNM effects on the $J/\psi$ production, let us
consider the $J/\psi$ production at CERN SPS experiments.
Fig.~\ref{Isayev:fig:3} shows the $J/\psi$ production cross
section, where $J/\psi$ are registered through their decay in the
dimuon channel, normalized to the cross section of the Drell-Yan
(DY) process~\cite{Isayev:S}. The nuclear thickness parameter $L$
parameterizes the number of participating nucleons and increases
with centrality.  At SPS energies,  DY process is not modified by
the medium and its production cross section scales linearly with
the number of binary NN collisions. Hence, this process can be
used as unsuppressed reference process for the $J/\psi$
production. The current interpretation of the obtained results is
the following. The effects of cold nuclear matter are quantified
in proton-nucleus collisions,  where the energy density is not
enough to reach the critical value for the formation of QGP. There
is no anomalous suppression, beyond CNM effects, of the $J/\psi$
yield in the collisions of light ions of sulfur ($A=32$) with the
heavy ions of uranium ($A=238$). However, there is the anomalous
suppression beyond the CNM effects in the central collisions of
lead nuclei when the number of participating nucleons is the
largest, and, to a certain extent, in the central collisions of
indium nuclei ($A=115$)~\cite{Isayev:Arn}. This could be
considered as the signature of melting the charmonium states in
quark-gluon plasma. At this point, it is also worth to note that
some level of the $J/\psi$ suppression could originate with the
reduced feed-down to $J/\psi$ from excited charmonium states
($\psi',\chi_c$), which melt just above the QGP transition
temperature. One needs further to investigate  what is really
suppressed, the directly produced $J/\psi$s or the $J/\psi$s
coming from the heavier  charmonium states.

Let us consider the $J/\psi$ production at RHIC experiments, when
the nucleon-nucleon center-of-mass energy for collisions of the
heaviest ions increases by a factor 10 compared to the CERN SPS
experiments. The $J/\psi$ suppression can be characterized by a
ratio called the nuclear modification factor
\begin{equation}R_{AB}(p_{\,T},y)=
\frac{d^2N_{J/\psi}^{AB}/dp_{\,T}dy}{N_{coll}d^2N_{J/\psi}^{pp}/dp_{\,T}dy},\end{equation}
 obtained by normalizing the $J/\psi$ yield in nucleus-nucleus
collision  by the $J/\psi$ yield in $p+p$ collision at the same
energy per nucleon pair times the average number  of binary
inelastic NN collisions.  This ratio characterizes the impact of
the medium on the particle spectrum. If heavy ion collision is a
superposition of independent $N_{coll}$ inelastic NN collisions,
then $R_{AB}=1$, whereas $R_{AB}<1$ ($R_{AB}>1$) corresponds to
the case of the $J/\psi$ suppression (enhancement). As we
discussed already, at first, it is necessary to clarify the role
of  CNM effects on $J/\psi$ production. At RHIC, CNM effects are
studied in collisions of light deuteron and heavy gold nuclei,
when the energy density reached in the collision is not enough for
the formation of QGP.   At RHIC energies, shadowing of partons is
important and in the model calculations is implemented in two
shadowing schemes for the nuclear parton distribution functions,
the EKS model~\cite{Isayev:EKS} and NDSG model~\cite{Isayev:NDSG}.
The $J/\psi$ break-up cross sections obtained for two shadowing
schemes from the best fit to data are $\sigma_{\rm
breakup}=2.8^{+2.3}_{-2.1}$ mb (EKS) and $\sigma_{\rm
breakup}=2.6^{+2.2}_{-2.6}$ mb (NDSG)~\cite{Isayev:Adare} (in
Ref.~\cite{Isayev:Adare}, the term "break-up cross section" is
used  instead of the term "absorption cross section"). Although
these values are consistent, within large uncertainties, with the
corresponding value obtained at CERN SPS, a recent
analysis~\cite{Isayev:L}  shows that, in fact, the level of
$J/\psi$ CNM break-up significantly decreases with the collision
energy.

\begin{figure}[tb]
\begin{center}
  \begin{tabular}{cc}
  \includegraphics[bb= 0 0 567 479,width=6.8cm]{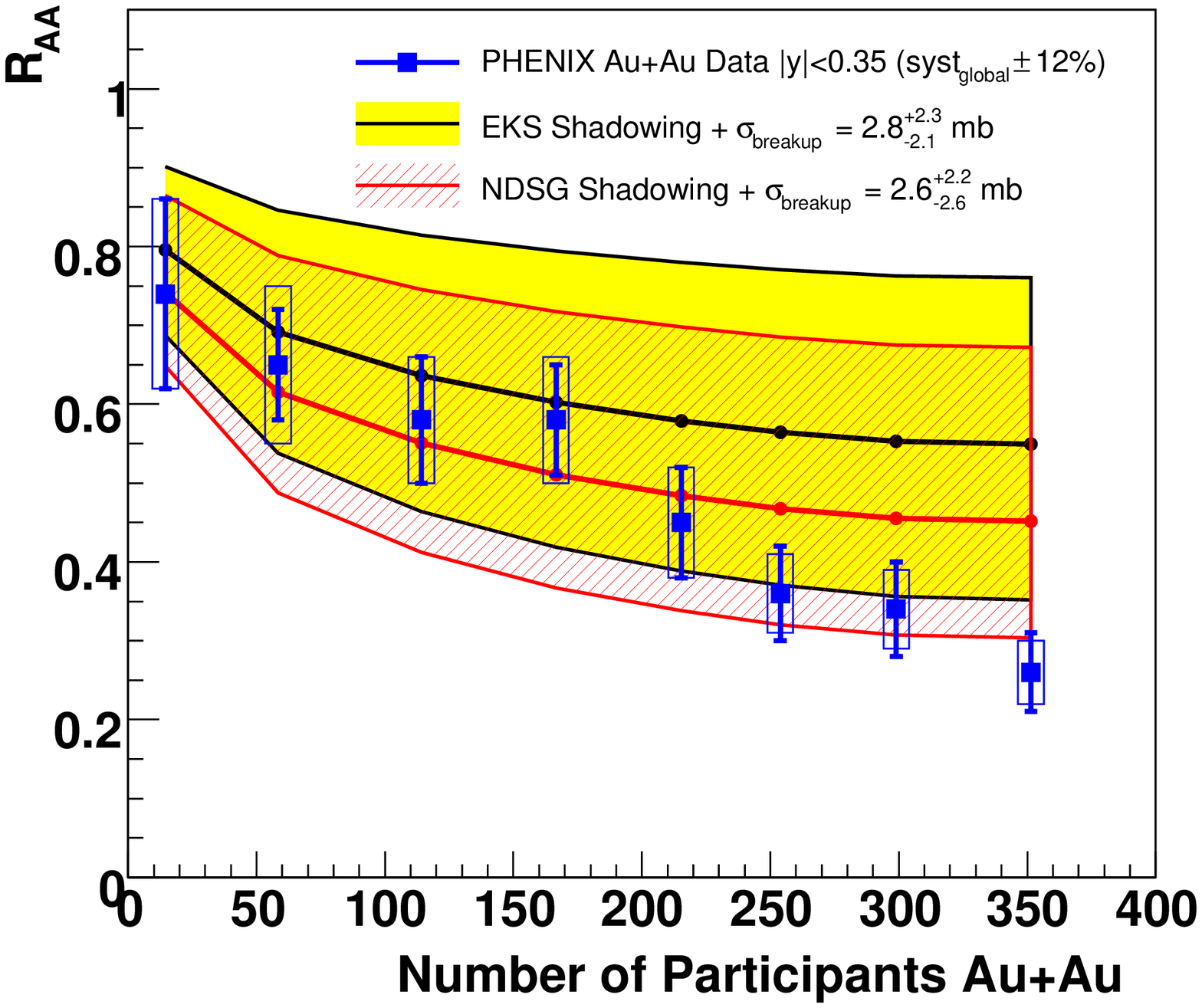} &
  \includegraphics[bb= 0 0 567 479,width=6.8cm]{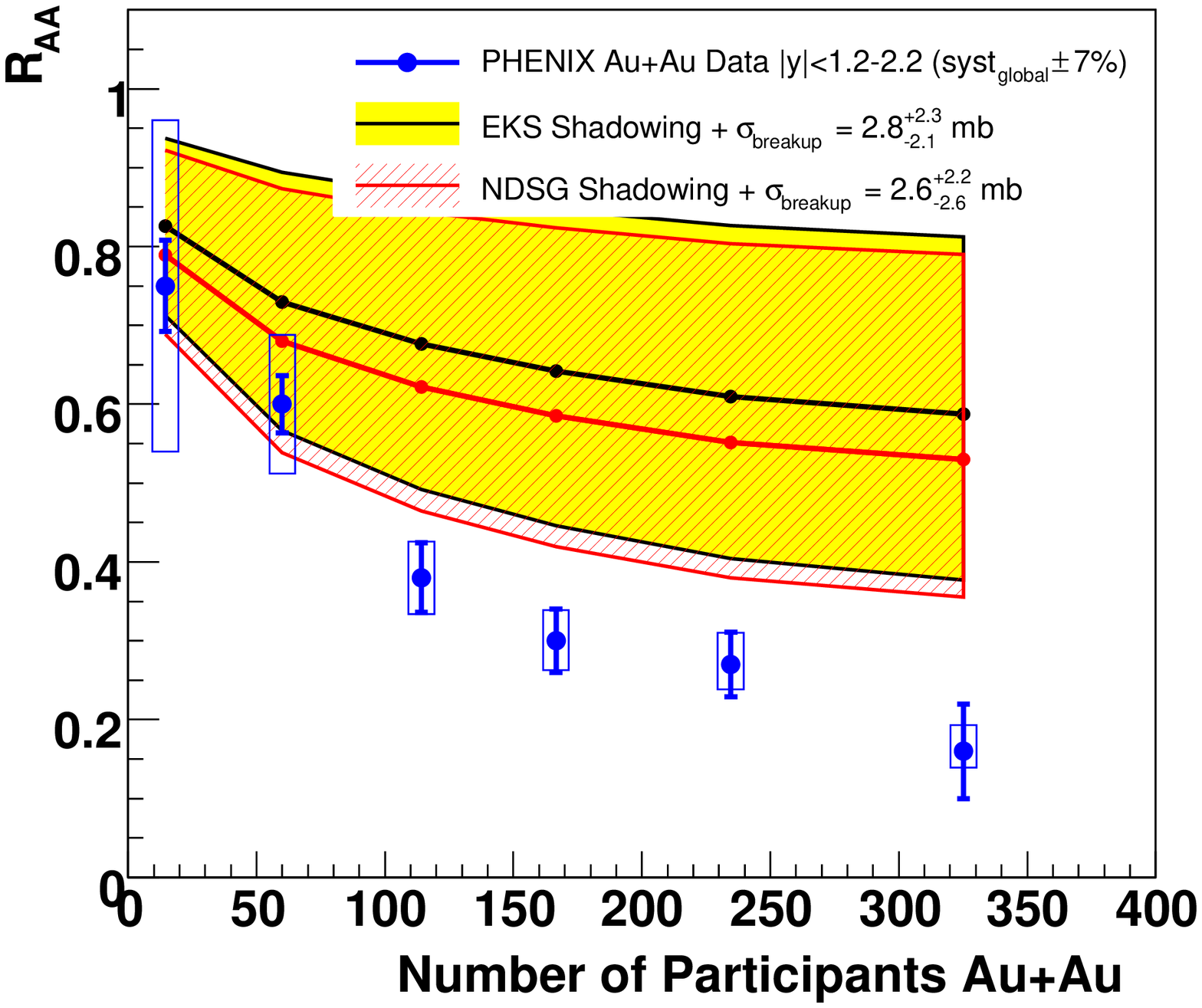}
  \end{tabular}
  \end{center}
\caption{Left: $J/\psi$'s $R_{AA}$ for $Au$-$Au$ collisions at
midrapidity compared to a band of theoretical curves for the
breakup values found to be consistent with the $d$-$Au$ data. Both
EKS and NDSG shadowing schemes are included. Right: Same as on the
left but at forward rapidity.} \label{Isayev:fig:4}
\end{figure}

Let us now consider charmonium production in heavy ion collisions
at RHIC. Fig.~\ref{Isayev:fig:4}  shows the $p_T$-integrated
$J/\psi$ nuclear modification factor obtained in Au-Au collisions
at RHIC/PHENIX experiment as a function of centrality,
parametrized by the number of participating nucleons at mid- (left
panel) and forward (right panel) rapidity~\cite{Isayev:A}. The
PHENIX data are shown by blue symbols. The $R_{AA}$ approaches
unity for the peripheral collisions (small $N_{part}$) and goes
down to approximately 0.2 at most central collisions (large
$N_{part}$). To see the level of the anomalous suppression beyond
the cold nuclear matter effects it is necessary to extrapolate the
CNM effects obtained in d-Au collisions to the Au-Au collisions
within the given shadowing scheme and the $J/\psi$ break-up cross
section. The results are shown by black and red curves with the
corresponding error bands. It is seen that $J/\psi$ production is
significantly suppressed beyond CNM effects at forward rapidity
(right) and suppression is less pronounced at midrapidity (left)
in most central Au-Au collisions.

\begin{figure}[tb]\begin{center}
\includegraphics[width=6.8cm]{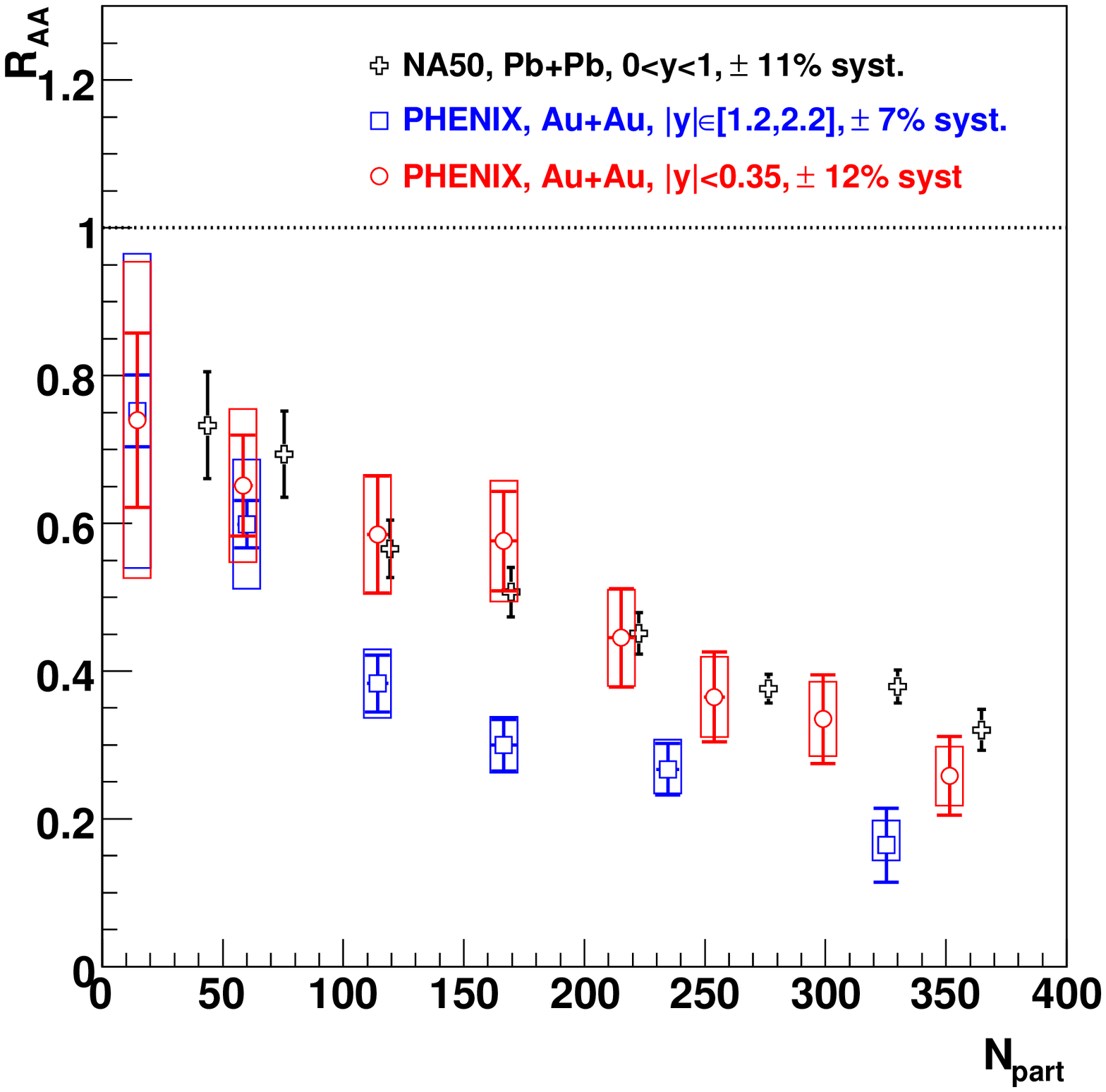}
\end{center}
\caption{$J/\psi$ nuclear modification factor for the most
energetic SPS (Pb-Pb) and RHIC (Au-Au) collisions, as a function
of the number of participants $N_{part}$. }
\label{Isayev:fig6}
\end{figure}

However, not all is still clear. Measurements of the $J/\psi$
suppression by PHENIX collaboration at RHIC lead to some
surprising features. Fig.~\ref{Isayev:fig6} shows compiled data
for the nuclear modification factor obtained in CERN SPS and RHIC
PHENIX experiments. There are two surprising results in these
measurements. First, the mid rapidity suppression in PHENIX (the
red boxes) is lower than the forward rapidity suppression (blue
boxes) despite the experimental evidence that the energy density
is higher at midrapidity than at forward rapidity, and, hence, one
could expect that at midrapidity the $J/\psi$s will be more
suppressed due to higher density of color charges. Secondly, the
nuclear modification factor $R_{AA}$ at midrapidity in PHENIX (red
boxes) and SPS (black crosses) are in agreement within error bars,
a surprising result considering that the energy density reached at
RHIC is larger than the one reached at SPS. This indicates that at
RHIC energies additional mechanisms countering the suppression,
could be operative.

Let us consider possible explanations of the above features. 1.
Regeneration of $J/\psi$s in the hot partonic phase from initially
uncorrelated $c$ and $\bar c$ quarks (quark coalescence model). If
to compare the $J/\psi$ suppression pattern at RHIC and SPS,
$J/\psi$s could be indeed more suppressed at RHIC than at SPS, but
then regenerated during (or at the boundary of) the hot partonic
phase from initially uncorrelated $c$ and $\bar c$ quarks. If to
compare the results at RHIC (midrapidity vs. forward rapidity), at
midrapidity, due to higher energy density, there are more $c$ and
$\bar c$ to regenerate than at forward rapidity that could explain
the stronger suppression at forward rapidity. Note that the total
number of initial $c\bar c$ pairs is larger than 10 in the most
central Au-Au collisions. Certainly, if regeneration is important
at the RHIC conditions, it will be even more important at the LHC
conditions where more than 100 $c\bar c$ pairs is expected to be
produced in the central Pb-Pb collisions. 2. $J/\psi$ production
could be more suppressed at forward rapidity due to the nuclear
shadowing effects. Standard gluon shadowing parametrizations do
not tend to produce such an effect but they are poorly constrained
by the data and further saturation effects are not
excluded~\cite{Isayev:dC}.

To show the complexity of the problem, let us consider some
theoretical models for the charmonium production, which quite
satisfactory describe the RHIC data but whose predictions for LHC
are drastically different. First, in the {\it statistical
hadronization model} (SHM)~\cite{Isayev:ABRS}, it is assumed that:
1. All heavy quarks (charm and bottom) are produced in primary
hard collisions and their total number stays constant until
hadronization. 2. Heavy quarks reach thermal equilibrium in the
QGP before the chemical freeze-out (hadronization). 3.~All
quarkonia are produced (nonperturbatively) through the statistical
coalescence of heavy quarks at hadronization. Multiplicities of
various hadrons are calculated with the grand canonical ensemble.
The generation of $J/\psi$ proceeds effectively if $c, \bar c$
quarks are free to travel over large distances implying
deconfinement.

\begin{figure}[tb]
\begin{center}
  \includegraphics[bb= 0 0 527 307,width=10cm]{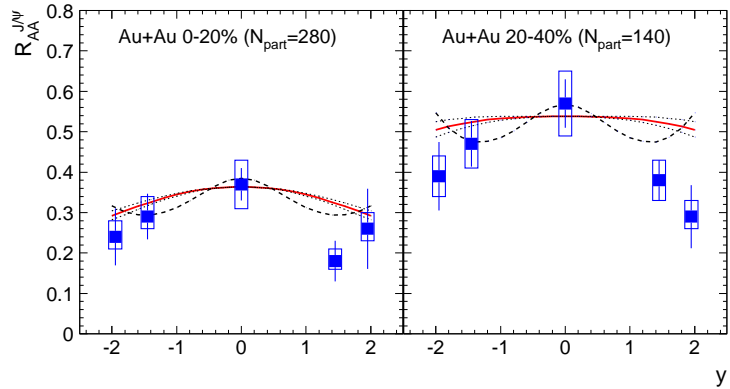}
  \end{center}
  \caption{Rapidity dependence of  $R_{AA}^{J/\psi}$
for two centrality classes in the statistical hadronization model.
The data from the PHENIX experiment (symbols with errors) are
compared to calculations (lines, see text).}\label{Isayev:fig:7}
\end{figure}

\begin{figure}[tb]
\begin{center}
 \includegraphics[bb= 0 0  507 507,width=0.42\textwidth]{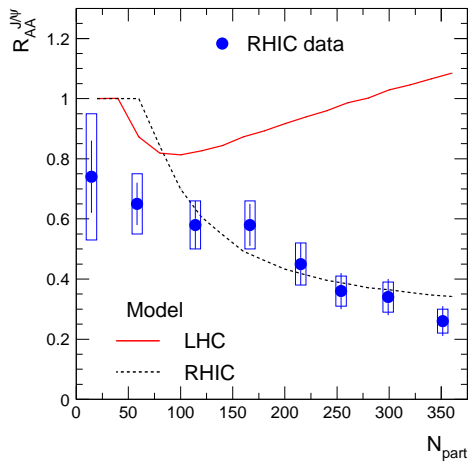}
   \end{center}
  \caption{Centrality dependence of the   $R_{AA}^{J/\psi}$
 at midrapidity, according to the statistical hadronization model.}\label{Isayev:fig:8}
\end{figure}

Fig.~\ref{Isayev:fig:7} shows the rapidity dependence of the
nuclear modification factor, obtained in this model and the
comparison with the rapidity dependence at PHENIX for two
centrality bins. Two theoretical curves correspond to two fitting
procedures, with one and two Gaussians of the $J/\psi$ data in
$pp$ collisions. In both cases, calculations reproduce rather well
(considering the systematic errors) the $R_{AA}$ data. The model
describes the larger suppression away from midrapidity. The
maximum of $R_{AA}$ at midrapidity in this model is due to the
enhanced generation of charmonium around midrapidity, determined
by the rapidity dependence of the charm production cross section.
The centrality dependence of $R_{AA}$ at $y=0$ is shown in
Fig.~\ref{Isayev:fig:8}. The model reproduces quite well the
decreasing trend with centrality seen in the RHIC data.
Fig.~\ref{Isayev:fig:8} also shows the prediction of the model for
the LHC. At much higher LHC energies, the charm production cross
section is expected to be larger by about an order of magnitude.
As a result, a totally opposite trend as a function of centrality
is predicted, with $R_{AA}$ exceeding unity for central
collisions.

\begin{figure}[tb]
\begin{center}
  \begin{tabular}{cc}
  \includegraphics[bb= 0 0 567 542,width=6.8cm]{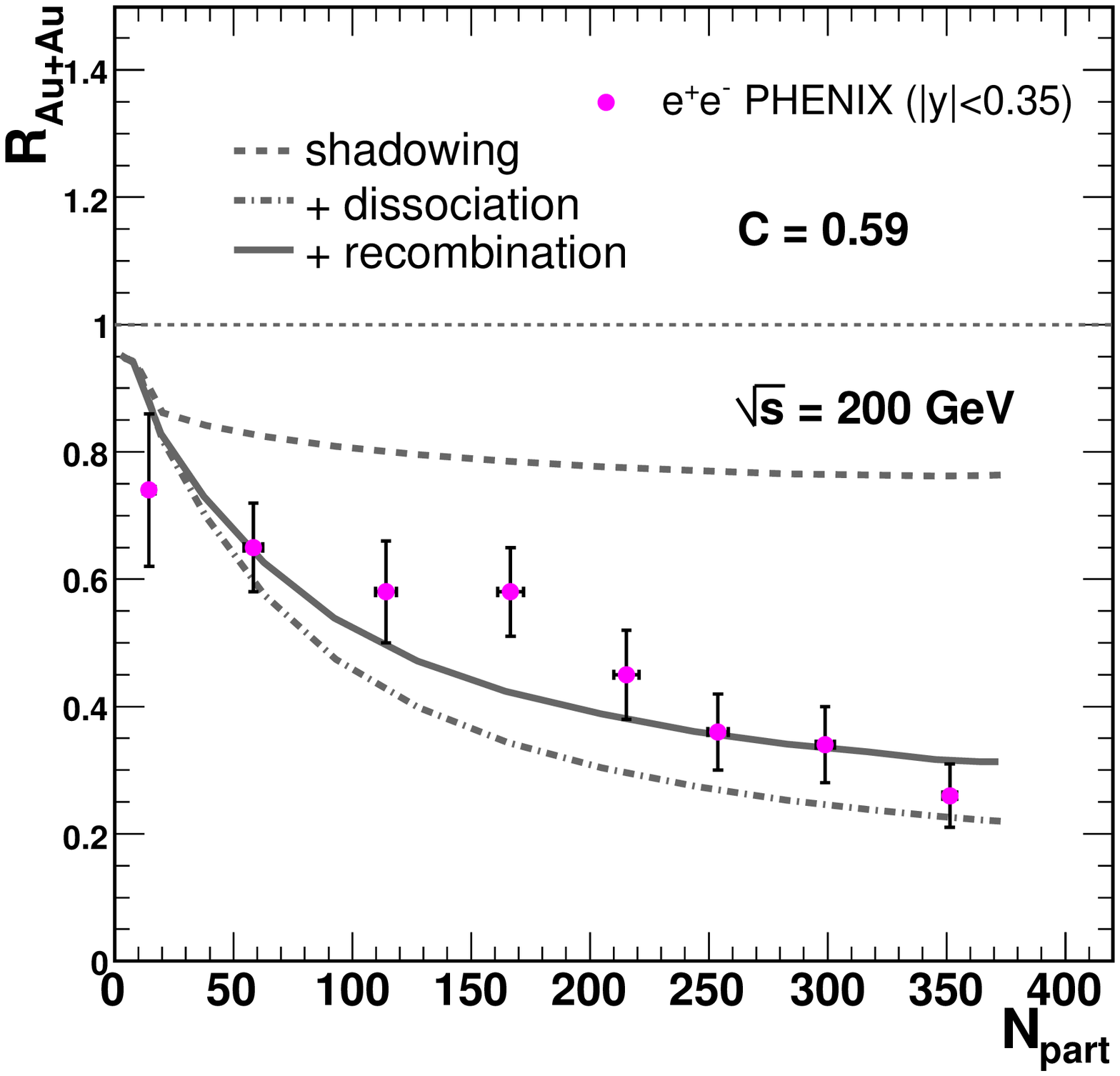} &
  \includegraphics[bb= 0 0 567 542,width=6.8cm]{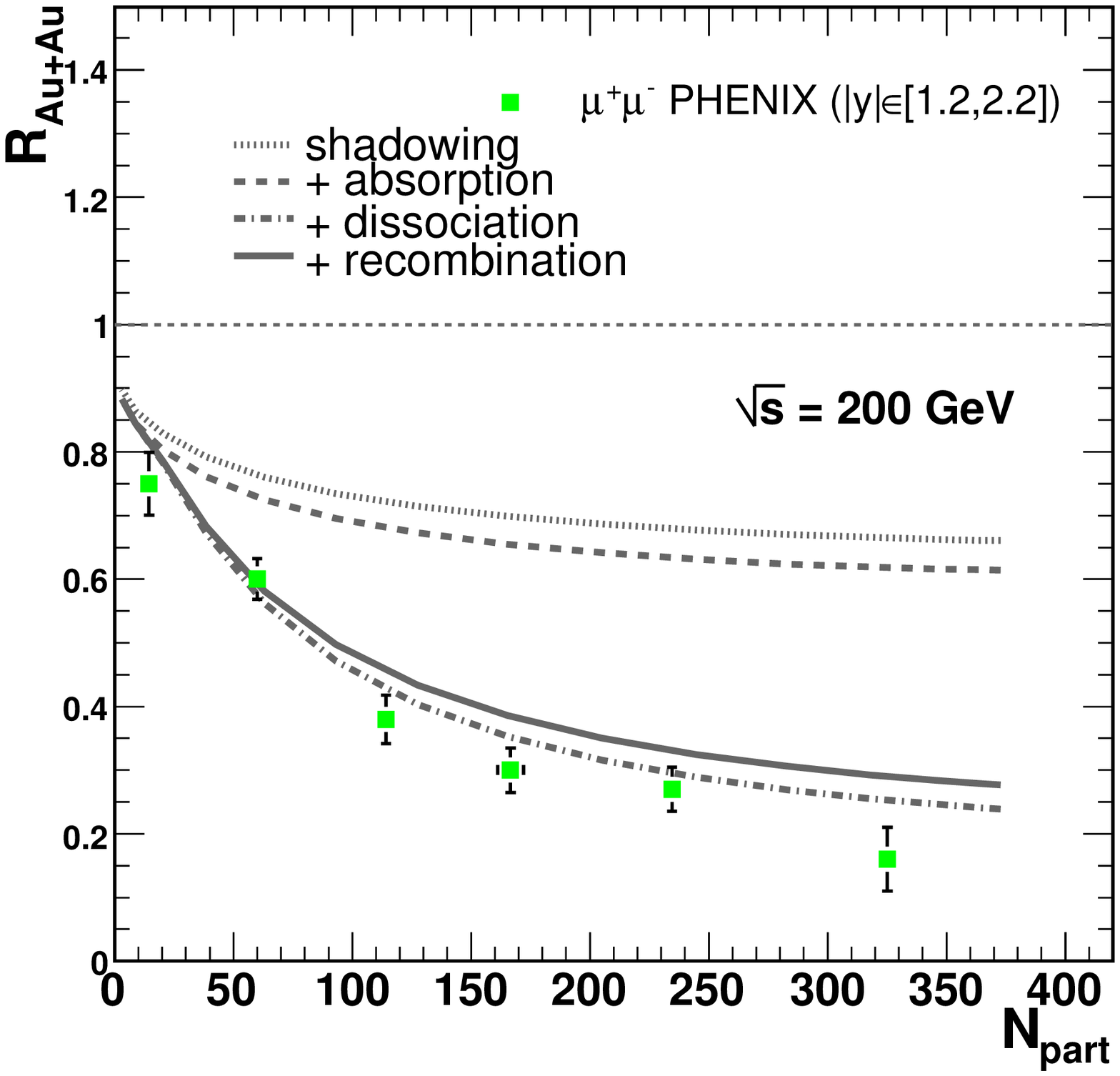}
  \end{tabular}
  \end{center}
\caption{Results for $J/\psi$ suppression in Au-Au collisions at
RHIC  at mid- (left panel), and at forward rapidities (right
panel) in the comovers interaction model. The solid curves are the
final results. The dashed-dotted ones are the results without
recombination ($C = 0$). The dashed line is the total
initial-state effect. The dotted line in the right panel is the
result of shadowing. In the left panel the last two lines
coincide. }
\label{Isayev:fig:9}
\end{figure}

\begin{figure}[tb]
\begin{center}
  \includegraphics[bb= 0 0 567 539,width=.4\textwidth]{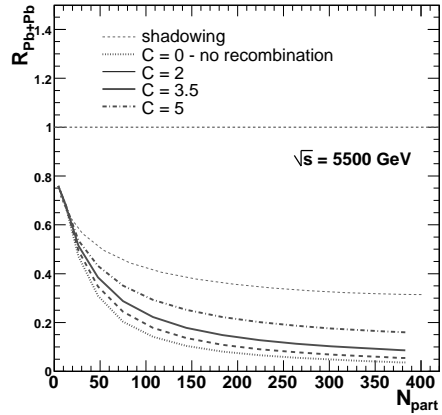}
  \end{center}
  \caption{Results for $J/\psi$ suppression in Pb+Pb at LHC ($\sqrt{s} = 5.5$ TeV) at
  midrapidity
  for different
values of the parameter $C$ in the comovers interaction model. The
upper line is the suppression due to initial-state effects
(shadowing).}\label{Isayev:fig:10}
\end{figure}

Let us consider {\it the comovers interaction model}
(CIM)~\cite{Isayev:CBFK}. This model does not assume the
deconfinement phase transition. Anomalous suppression of $J/\psi$
(beyond CNM effects) is the result of the final state interaction
of the $c \bar c$ pair with the dense medium produced in the
collision (comovers interaction). The model consistently treats
the initial and final state effects. The initial state effects
include: 1) nuclear absorption of the pre-resonant $c \bar c$
pairs by nucleons of the colliding nuclei, 2) consistent treatment
of nuclear shadowing for hard production of charmonium. The final
state effects include absorption of the $c \bar c$ pairs by the
dense medium created in the collision (interaction with comoving
hadrons produced in the collision). The model does not assume
thermodynamic equilibrium and, thus, does not use thermodynamic
concepts. The density of charmonium is governed by the
differential rate equation
\begin{equation}
\tau\frac{dn_{J/\psi}}{d\tau}=-\sigma_{co}[n_{co}(b,s,y)n_{J/\psi}(b,s,y)-n_c(b,s,y)
n_{\bar c}(b,s,y)],\label{Isayev:2}
\end{equation}
supposing a pure longitudinal expansion of the system and boost
invariance. In Eq.~(\ref{Isayev:2}), $n_{co}$ is the density of
comovers, which is found in the dual parton
model~\cite{Isayev:CSTT} together with the proper shadowing
correction, $\sigma_{co}$ is the cross section of $J/\psi$
dissociation due to interactions with comovers, taken such as to
reproduce the low energy SPS experimental data (with $\sigma_{co}
= 0.65$ mb). The first term on the right describes dissociation of
charmonium due to interaction with comovers. The second term
describes the recombination of charmonium and is proportional to
the product of densities of charm quarks and antiquarks. The
important feature of the CIM is that recombination of $c$-$\bar c$
quarks proceeds only locally, when the densities of quarks and
antiquarks are taken at the same transverse coordinate $s$. This
is different from the recombination in the SHM, where recombining
quarks can be separated by large distance that implies
deconfinement. The effective recombination cross section in the
CIM is equal to the dissociation cross section due to the detailed
balance. The results for the centrality dependence of the $R_{AA}$
are shown in Fig.~\ref{Isayev:fig:9}. At midrapidity, the
experimental data are well reproduced by full theoretical
calculations (solid curve) taking into account nuclear shadowing,
dissociation by comovers and recombination from charm quark and
antiquark pairs. At forward rapidity, the results also well agree
with data, in particular, the $J/\psi$ suppression at forward
rapidity is somewhat larger that the suppression at midrapidity.
 Fig.~\ref{Isayev:fig:10} shows the
predictions of the model for LHC. The parameter $C$ encodes the
recombination from $c$-$\bar c$ pairs and is vanishing in the
absence of recombination. Although the density of charm grows
substantially from RHIC to LHC, the combined effect of
initial-state shadowing, absorption and comovers dissociation
overcomes the effect of parton recombination. This is in sharp
contrast with the predictions of the statistical hadronization
model where a strong enhancement of the $J/\psi$ yield with
increasing centrality was predicted.

Thus, the $J/\psi$ suppression is an important characteristic to
search for QGP. But if $J/\psi$ anomalous suppression (beyond CNM
effects) was observed in HIC, there are a few competing mechanisms
to explain that: 1. Charmonium is dissociated due to the genuine
color screening in the deconfined medium. 2. Charmonium is
dissociated through interactions with comoving hadrons in the
medium formed in HIC. How to differ these mechanisms? In comovers
interaction model, the anomalous suppression sets in smoothly from
peripheral to central collisions rather than in a sudden way when
the corresponding dissociation temperature in the deconfined
medium is reached. At SPS, current experimental errors still do
not allow to disentangle these two mechanisms. However, even if
the color screening mechanism is dominating, it is unclear what is
really melted, directly produced $J/\psi$s or originating from the
feed-down of less bound charmonium states, $\chi_c$ ($\rightarrow
J/\psi+X$), $\psi'$ ($\rightarrow J/\psi+X$), which have lower
dissociation temperatures.  At RHIC and LHC conditions, the
feed-down from B-meson decays becomes also important.
Recombination enhances $J/\psi$ production and much complicates
the picture but its effect may be different depending on whether
the deconfinement phase transition happened or not. It is even
possible that after all $J/\psi$s were melted in QGP they can be
statistically regenerated at hadronization. True operative
mechanisms of $J/\psi$ production can be established only after
studying all important dependences (from centrality, rapidity,
collision energy $\sqrt{s}$,...) of all relevant observables with
sufficient accuracy.

So far we considered the suppression (or enhancement) patterns for
the $J/\psi$ production. Now let us briefly consider the
perspective for bottomonia. One could expect that bottomonium
production might be easier to understand than charmonium
production due to the following reasons. 1. Since less than one
$b\bar b$ pair is produced in one central Au-Au collision, the
regeneration is negligible at the conditions of RHIC. Besides,
only about 5 $b\bar b$ pairs are expected to be produced in a
single central Pb-Pb collision at LHC. Hence, regeneration should
play much less role in the beauty sector than in the charm sector.
2. Having higher masses, bottomonia originate from higher momentum
partons and will less suffer from shadowing effects. 3. The
absorption cross section for $\Upsilon$ is by $40-50\,\%$
 smaller than the corresponding cross
section for $J/\psi$ and $\psi'$.

These features should ease the separation of the anomalous
suppression in the $\Upsilon$'s family.

\section{Heavy Flavor Probes of QGP: Open Heavy Flavor Mesons}

What qualitative effects could one expect to obtain when probing
the dense matter by heavy quarks (charm or bottom)? As  well known
from electrodynamics, the bremsstrahlung off an accelerated heavy
quark $Q$ is suppressed by the large power of its mass $\sim
(m_q/m_Q)^4$
 as compared to light quarks. Therefore,
gluon radiation off heavy quarks (i.e., radiative energy loss) is
much suppressed relative to light quarks. Consequently, one could
expect a decrease of high $p_T$ suppression and of the elliptic
flow coefficient $v_2$~\cite{Isayev:I} from light to charm to
bottom quarks. Or, that the energy loss and coupling to matter of
heavy quarks is smaller than for light quarks as well as that the
thermalization time for heavy quarks is longer than for light
quarks. Due to the above features, one should observe a pattern of
gradually increasing $R_{AA}$ when going from the mostly
gluon-originated light-flavor hadrons ($h^\pm$ and $\pi^0$) to $D$
  to $B$ mesons: $R_{AA}^h\lesssim R_{AA}^D\lesssim
R_{AA}^B$~\cite{Isayev:Ar} (gluons lose more energy than quarks
since gluons have a higher color charge). The enhancement above
the unity of the heavy-to-light ratio
$R_{AA}^{D/h}=R_{AA}^D/R_{AA}^h$ probes the color charge
dependence of the parton energy loss while the ratio
$R_{AA}^{B/D}=R_{AA}^B/R_{AA}^D$ probes the mass dependence of the
parton energy loss.

\begin{figure}[tb]
\begin{center}
  \includegraphics[bb= 10 15 550 545,width=.4\textwidth]{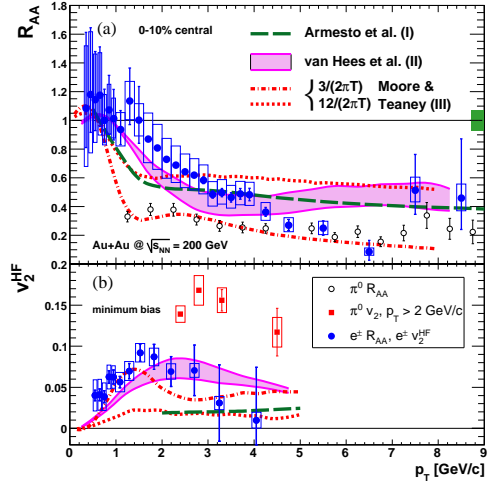}
  \end{center}
  \caption{Nuclear modification factor (upper panel, central Au-Au) and elliptic flow
(lower panel, minimum-bias Au-Au) of non-photonic electrons at
RHIC, compared to theory. The band corresponds to the Langevin
simulations based on an expanding fireball with effective heavy
quark resonance interactions~[22].} \label{Isayev:fig:11}
\end{figure}

Let us now consider what the experiment tells us about the open
heavy flavor $p_T$ suppression and elliptic flow. As mentioned
earlier, at RHIC, open heavy flavors are studied through
measurements with the electrons and positrons originating from the
semileptonic decays of $D$- and $B$-mesons.
Fig.~\ref{Isayev:fig:11} shows the nuclear modification factor and
elliptic flow coefficient for HF electrons as functions of $p_T$,
obtained in central collisions of gold nuclei at RHIC (closed
circles)~\cite{Isayev:A2}-\cite{Isayev:R}. In contrast to the
above expectations, the results for $R_{AA}^{HF}$ show a strong
suppression of HF decay electrons at $p_T>2$ GeV/c, approaching at
high $p_T$ the level of suppression for $\pi^0$. This evidences
that produced medium is quite dense for heavy quarks to lose
energy as efficiently as light quarks do. The measurement of
elliptic flow gives rather large value for $v_2^{HF}$. This means
that HF electrons are involved in a collective motion being
indicative of the collective flow of their parent particles as
well.

In Fig.~\ref{Isayev:fig:11}, the results of some model
considerations are also shown. The best description is provided by
the model assuming the Brownian motion of heavy quarks within the
framework of Langevin dynamics~\cite{Isayev:HGR}. Let us consider
the basic assumptions of this theory. Firstly, the thermal heavy
quark momentum $p^2\sim mT$ (in nonrelativistic approximation) is
much larger than the typical momentum transfer $Q^2\sim T^2$ from
a thermal medium to a heavy quark. Hence, the motion of a heavy
quark in the QGP can be represented as the Brownian motion, which
can be described using the  Langevin equation. Secondly, heavy
quark loses its energy in elastic scattering processes with light
partons. Besides, as evidenced from calculations of heavy and
light meson correlators within LQCD, in QGP the $D$- and $B$-meson
like resonant states exist up to $T<2T_c$. Rescattering on these
resonant states plays an important role in thermalizing heavy
quarks. Thirdly, in order to get the spectrum of HF electrons,
$c$- and $b$-quarks are to be hadronized into $D$- and $B$-mesons
via quark coalescence (at low $p_T$) and fragmentation (at high
$p_T$).

The analysis shows  that resonance scattering decreases nuclear
modification factor $R_{AA}^{HF}$ and increases azimuthal
asymmetry $v^{HF}_2$. Heavy-light quark coalescence in subsequent
hadronization significantly amplifies $v^{HF}_2$ and increases
$R_{AA}^{HF}$, especially in the $p_T\simeq2$ GeV/c region. The
contribution from B-mesons to $R_{AA}^{HF}$ and $v^{HF}_2$ is
estimated providing full calculations with $c$+$b$ quarks and with
only $c$ quarks. The result is that the B-meson contribution
increases $R_{AA}^{HF}$ and decreases $v^{HF}_2$, and becomes
important above $p_T\simeq3$ GeV/c. Thus, one can conclude, that
the combined effects of coalescence of heavy quarks $Q$ with light
quarks $q$, and of the resonant heavy-quark interactions are
essential in generating strong elliptic flow $v^{HF}_2$ of up to
$10\%$, together with strong suppression of heavy flavor electrons
with $R^{HF}_{AA}$ about 0.5.



\end{document}